\documentclass[iop,twocolumn,tighetn,10pt,apj]{emulateapj}
\usepackage{multirow}
\shorttitle{Accretion Powered Stellar Winds}
\shortauthors{Zanni \& Ferreira}

\begin{document}

\title{Observational limits on the spin-down torque of Accretion Powered Stellar Winds}

\author{Claudio Zanni}
\affil{INAF - Osservatorio Astronomico di Torino \\
        Strada Osservatorio 20, 10025, Pino Torinese, Italy}
\email{zanni@oato.inaf.it}    
\and

\author{Jonathan Ferreira}
\affil{UJF-Grenoble 1/CNRS-INSU \\ 
         Institut de Plan\'etologie et d'Astrophysique de Grenoble (IPAG) UMR 5274 \\
          Grenoble, F-38041, France}
\email{Jonathan.Ferreira@obs.ujf-grenoble.fr}

\begin{abstract}
The rotation period of classical T~Tauri stars (CTTS) represents a longstanding puzzle.
While young low-mass stars show a wide range of rotation periods, many CTTS are 
slow rotators, spinning at a small fraction of break-up, and their rotation period does
not seem to shorten, despite the fact that they are actively accreting and contracting.
\cite{matt05b} proposed that the spin-down torque of a stellar wind 
powered by a fraction of the accretion energy would be strong enough to balance the spin-up torque due to accretion.
Since this model establishes a direct relation between accretion and ejection, the observable stellar parameters
(mass, radius, rotation period, magnetic field) and the accretion diagnostics (accretion shock luminosity), can be used
to constraint the wind characteristics. In particular, since the accretion energy powers both the stellar wind
and the shock emission, we show in this letter how the accretion shock luminosity $L_\mathrm{UV}$ can provide
upper limits to the spin-down efficiency of the stellar wind. 
It is found that luminous sources  with $L_\mathrm{UV} \geq 0.1 L_\odot$ and typical dipolar field components $< 1$ kG 
do not allow spin equilibrium solutions. 
Lower luminosity stars ($L_\mathrm{UV} \ll 0.1 L_\odot$) are compatible with a zero-torque condition, but the 
corresponding stellar winds are still very demanding in terms of mass and energy flux. 
We therefore conclude that accretion powered stellar winds are unlikely to be the sole mechanism to provide an efficient 
spin-down torque for accreting classical T~Tauri stars.
\end{abstract}


\section{Introduction}

Classical T~Tauri stars (CTTS) are known to be magnetically active protostars showing clear observational signatures of accretion 
(circumstellar disks) and ejection (jets and outflows). The stellar magnetic fields measured by spectropolarimetric observations 
\citep[up to a few kG,][]{JK07} can deeply affect the dynamics of the circumstellar region: truncating the disk and channeling the 
accretion flow along the magnetic surfaces down to the stellar surface; providing an acceleration mechanism 
for different types of outflows, stellar winds along the opened magnetospheric fieldlines \citep{Matt08a} and ejections associated 
with the magnetic star-disk interaction \citep{Shu94,Ferreira00,Romanova09,Zanni09}.

Their spin represents a controversial issue. While a wide range of rotation periods is observed among low-mass
young stars \citep[0.2 up to 20 days,][]{Irwin09}, around half of them slowly rotate, much below the break-up limit.
Besides, many slow rotators show clear accretion signatures \citep{Herbst07}, as in the case of CTTS, which have an 
average rotation period around  $\sim8$ days, corresponding to $\sim10\%$ of their break-up speed \citep{Bouvier93}.
Since CTTS are accreting mass and angular momentum from the surrounding accretion disk and they are still contracting,
they would be expected to noticeably spin-up in a few million years: conversely, there are indications 
that solar-mass slow rotators are prone to keep their rotation period constant for $\sim5$ Myr \citep{Irwin09}. 
Therefore, some mechanism must act to efficiently remove angular momentum from these slowly rotating stars.

Grounded on models originally developed to explain the period changes of pulsars \citep{GL79}, one of the most widespread scenarios foresees that
a significant spin-down torque is provided along the magnetospheric fieldlines connecting the star and the disk region rotating slower than the star \citep{ko91}.
On the other hand, both analytical \citep{matt05a} and numerical models \citep{ZF09} have questioned the efficiency of this mechanism, 
due to the limited extent of the connected region and the weakness of the magnetic connection. 

\begin{deluxetable*}{c l c c c c c c c c}
\tablecaption{Star sample\label{table:sample}}
\tablewidth{0pt}
\tablehead{
\multicolumn{2}{c}{Object} & \colhead{$M_\star$} &  \colhead{$R_\star$} &  \colhead{$B_\star$} & \colhead{$P_\star$} &  
\colhead{$\delta$} & \colhead{$L_\mathrm{UV}$} & \colhead{$\dot{M}_\mathrm{obs}$} & Ref. \\
\multicolumn{2}{c}{Name} & \colhead{[$M_\odot$]} & \colhead{[$R_\odot$]} &      \colhead{[G]}     &  \colhead{[days]}      &     
                              & \colhead{[$L_\odot$]} &  \colhead{[$M_\odot \;  \mathrm{\mbox{yr}}^{-1}$]}  & \\
}
\startdata
\multirow{2}{*}{BP~Tau} & (a)   & \multirow{2}{*}{0.7} & \multirow{2}{*}{1.95} & \multirow{2}{*}{600} & \multirow{2}{*}{7.6}  & \multirow{2}{*}{0.05} & 0.179 & $2 \times 10^{-8}$ & 1,3 \\
                                           & (b)   &                                    &                                      &                                     &                                      &                                       & 0.023 & $2.5 \times 10^{-9}$ & 3,5 \\
\multirow{2}{*}{V2129~Oph} &  (a)  & \multirow{2}{*}{1.35} & 2.4  & 175 & \multirow{2}{*}{6.5} & 0.06 &  0.143   & $                  10^{-8}$ & 2 \\
                                                  & (b)   &                                       & 2.1  & 450 &                                    & 0.05 &  0.01 &  $6.3 \times 10^{-10}$ & 6 \\
CV~Cha   &   & 2.0  & 2.5 & 300 & 4.4  & 0.07 &   0.61 & $3 \times 10^{-8}$ & 4\\
CR~Cha   &   & 1.9  & 2.5 & 200 & 2.3  & 0.14 & 0.02 &       $10^{-9}$  &  4\\
\multirow{2}{*}{AA~Tau} & (a) & \multirow{2}{*}{0.7} & \multirow{2}{*}{2} & \multirow{2}{*}{1500} & \multirow{2}{*}{8.2} & \multirow{2}{*}{0.05} & 0.025 & $2.8 \times 10^{-9}$ & 1,5 \\
                                          & (b)  &                                    &                                 &                                       &                                    &                                       & 0.006 & $6.3 \times 10^{-10}$ & 5 \\ 
\enddata
\tablerefs{[1] \citet{Gullbring98}, [2] \citet{Donati07}, [3] \citet{Donati08}, [4] \citet{Hussain09}, [5] \citet{Donati10a}, [6] \citet{Donati10b}}
\end{deluxetable*}

\citet{matt05b} have therefore proposed that stellar winds could efficiently remove angular momentum directly from the star along the opened 
fieldlines of the magnetosphere. Besides, they suggested that these outflows could derive their energy directly from the accretion power.
This would be also consistent with the fact that accreting CTTS seem to have on average longer rotation periods than their non-accreting
counterparts (weak-lined T~Tauri stars, WTTS), indicating a connection between spin-down and accretion \citep{Lamm05}.
It is commonly assumed that the accretion power is liberated in a shock due to the impact of the accretion streams with the stellar surface.
While a fraction of the accretion energy can be converted \citep[e.g. into Alfv\'en waves,][]{Sch88} and possibly injected into the wind, 
the emission of the shocked material can explain the observed optical excess and UV continuum \citep{Gull00}. 
Observations of the accretion shock luminosity can be therefore used to constrain the accretion energy which is available to power the stellar wind. 

In this letter we try to estimate the spin-down efficiency and the energy requirement of accretion powered stellar winds (APSW) compatible
with measurements of magnetic fields and accretion luminosities of several CTTS. In Section \ref{sec:model} we describe a simple analytical
APSW model and we apply it to a specific CTTS example in Section \ref{sec:BP}. We determine the stellar parameters which are compatible with
a spin equilibrium situation in Section \ref{sec:spin} and we summarize our conclusions in Section \ref{sec:dicsussion}.

\section{The model}
\label{sec:model}

In our analysis we will assume that, even in the case of a multipolar, complex stellar field, the dipolar component controls
the dynamics of both accretion (i.e. the disk truncation radius) and ejection (i.e. the wind magnetic lever arm). 
This assumption is somewhat supported by the results of \citet{Matt08a}: in the case of a multipolar field (quadrupolar in their case) 
with no dipolar component, the stellar wind torque is strongly suppressed. The same assumption 
is done evaluating the disk truncation radius: the dipolar component is the one which can affect the dynamics on a larger radial extent.

We report here the relevant equations to describe the dynamical properties of an APSW.
The spin-down torque exerted by a stellar wind characterized by a mass outflow rate 
$\dot{M}_\mathrm{wind}$ can be defined as $\dot{J}_\mathrm{wind} = \dot{M}_\mathrm{wind} \overline{r}_A^2 
\Omega_\star$, where $\Omega_\star=2\pi/P_\star$ is the angular speed of the star, $P_\star$ its rotation period 
and $\overline{r}_A$ is the wind average magnetic lever arm. For $\overline{r}_A$ we assume the \citet{Matt08a} approximation: 
\begin{equation}
\overline{r}_A = K\left(\frac{B_\star^2R_\star^2}{\dot{M}_\mathrm{wind}v_\mathrm{esc}}\right)^mR_\star\;, 
\label{eq:lever}
\end{equation}
where $K$ = 2.11 and $m$ = 0.223, obtained for a stellar wind flowing along the opened field lines of a dipolar 
magnetosphere. In Eq.~(\ref{eq:lever}) $M_\star$ and  $R_\star$ are the stellar mass and radius respectively, 
$B_\star$ is the intensity of the dipolar component of the magnetosphere at the stellar equator and 
$v_\mathrm{esc}=\sqrt{2GM_\star/R_\star}$ is the escape speed. 
Assuming a Keplerian disk rotation, we write the spin-up accretion torque as $\dot{J}_\mathrm{acc}=\dot{M}_\mathrm{acc} 
\sqrt{GM_\star R_\mathrm{t}}$, where $\dot{M}_\mathrm{acc}$ is the disk accretion rate and $R_\mathrm{t}$ is the disk truncation radius.
We assume that the radius at which the disk gets truncated by the dipolar component of the magnetosphere is proportional
to the Alfv\'en radius: $R_\mathrm{t}=C\Psi^{2/7}R_\star$, where $\Psi=B_\star^2R_\star^2/\dot{M}_\mathrm{acc}v_\mathrm{esc}$ 
is a dimensionless magnetization parameter. Different theoretical works limit the multiplying factor in the 
range $C\sim0.5-1$ \citep{Bessolaz08}: we assume an average value $C=0.75$. 
Combining all the previous expressions, the wind outflow rate necessary to extract a fraction 
$f_\mathrm{J}=\dot{J}_\mathrm{wind}/\dot{J}_\mathrm{acc}$ of the accretion torque is:
\begin{equation}
\dot{M}_\mathrm{wind}=\dot{M}_\mathrm{acc}\left(\frac{f_\mathrm{J}C^{1/2}\Psi^{1/7-2m}}{K^2\delta}\right)^\frac{1}{1-2m} \; ,
\label{eq:ejeff}
\end{equation}
where $\delta=\Omega_\star/\sqrt{GM_\star/R^3_\star}$  is the normalized stellar spin.
An analogous equation has been derived by \citet[][Eq. 17]{Matt08b}, assuming $f_\mathrm{J}=1$ and omitting the $C$ factor. 
A value $f_\mathrm{J}=1$ corresponds to a spin equilibrium situation, while $\delta=1$ represents a star rotating at break-up speed.
Employing Eq. (\ref{eq:ejeff}), we define the mass ejection efficiency as $f_\mathrm{M}=\dot{M}_\mathrm{wind}/\dot{M}_\mathrm{acc}$.

\begin{figure*}[t!]
   \centering
   \includegraphics[width=0.49\textwidth]{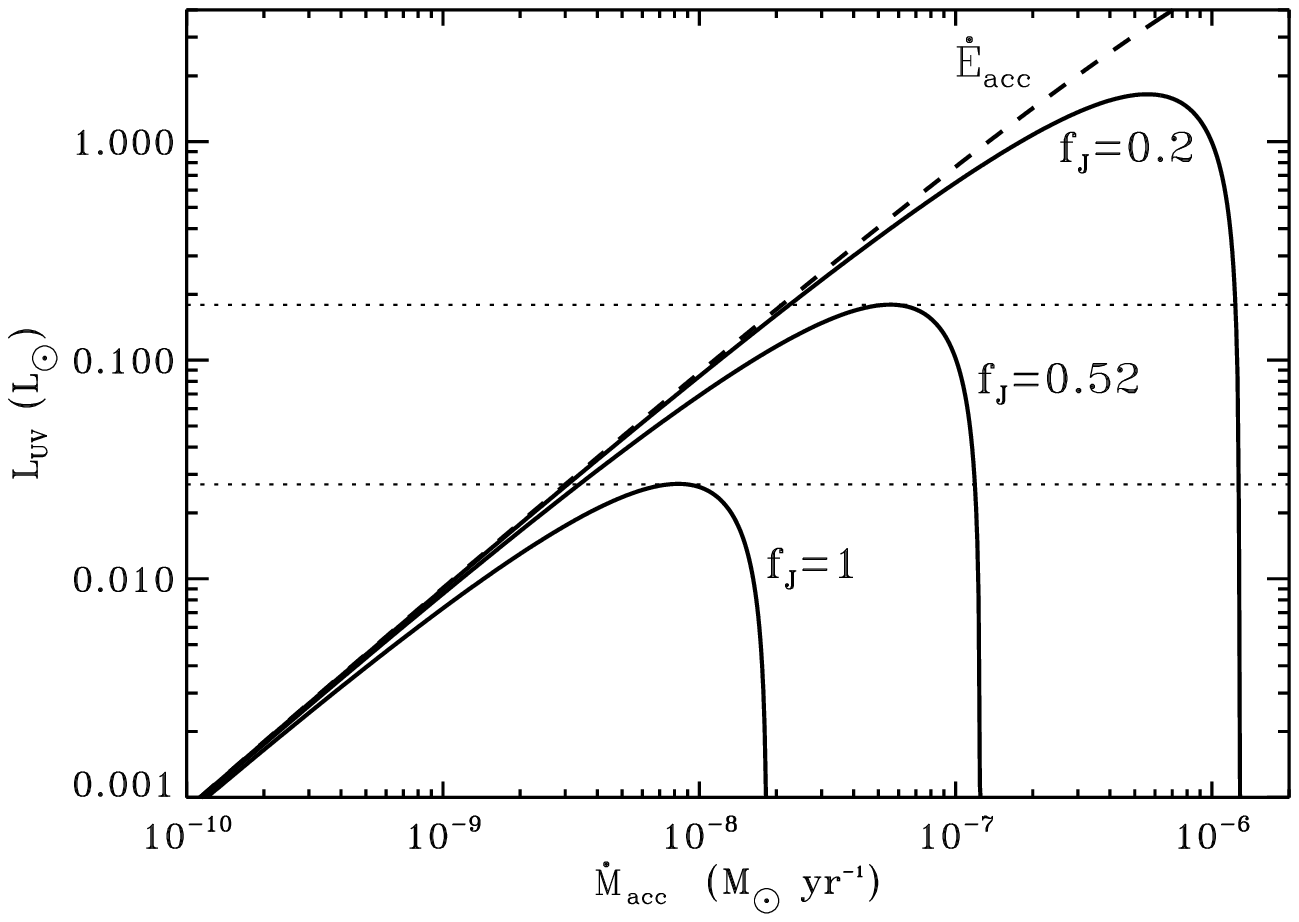}
   \includegraphics[width=0.49\textwidth]{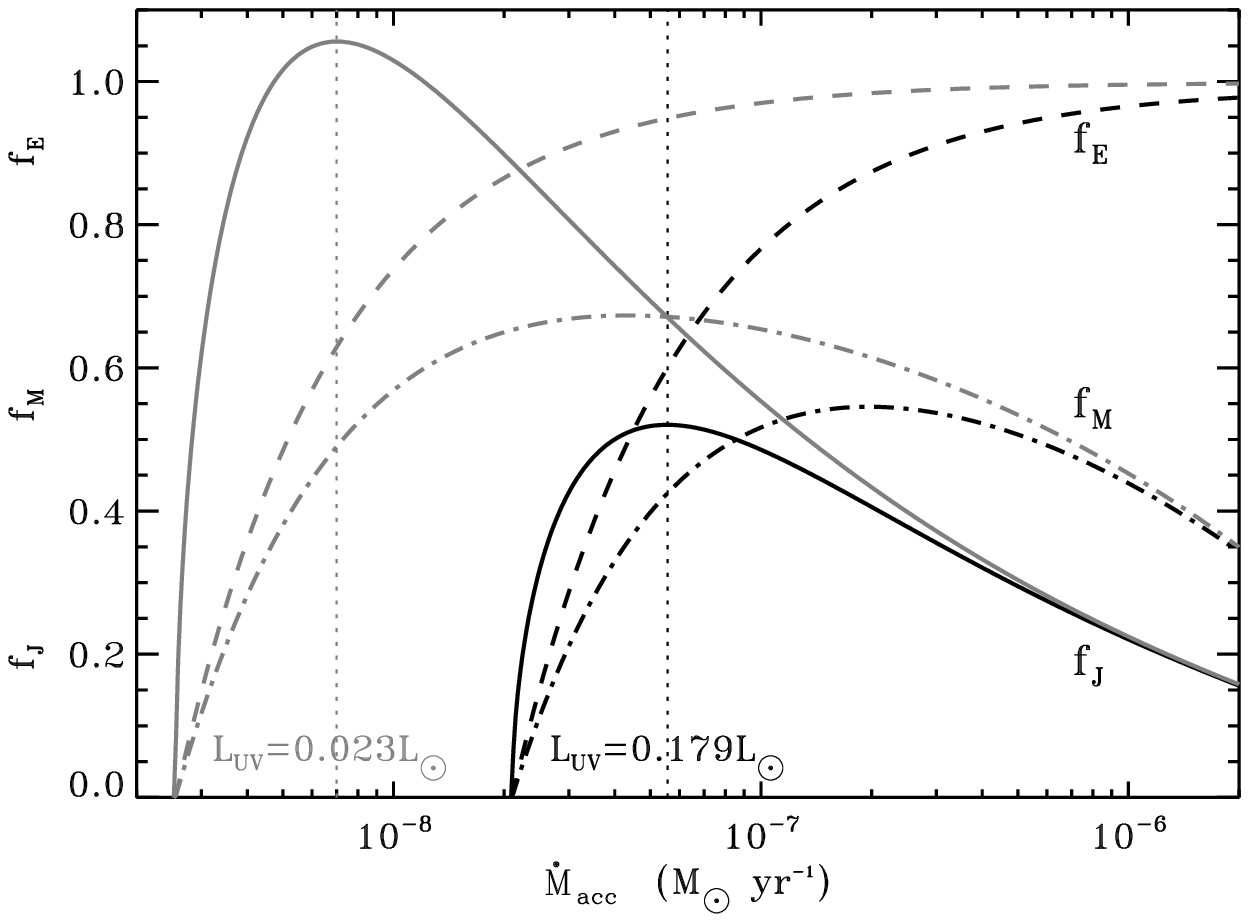}
  \caption{The BP Tau case. {\it Left panel}: accretion shock luminosity $L_\mathrm{UV}$ as a function of the accretion rate $\dot{M}_\mathrm{acc}$ 
                  for different values of the $f_\mathrm{J}$ parameter (solid lines). The dashed line represents the total available accretion power $\dot{E}_\mathrm{acc}$. 
                  The dotted lines represent the maximum accretion luminosities compatible with a given $f_\mathrm{J}$ value.
                  {\it Right panel}:  properties of the APSW as a function of the accretion rate for two possible accretion luminosities (grey and black lines);
                  $f_\mathrm{J}$ (solid lines), $f_\mathrm{M}$ (dot-dashed lines), $f_\mathrm{E}$ (dashed lines).}
  \label{fig:bptau}
\end{figure*}

How much energy is necessary to drive such a stellar wind? Since CTTS rotate well below their break-up speed ($\delta\ll1$), 
magneto-centrifugal processes are not efficient enough to accelerate the wind at the stellar surface and an extra energy input is required
to give to the wind the initial drive. This energy input corresponds roughly to a specific energy of the order of the potential gravitational energy:
$\dot{E}_\mathrm{wind}=1/2\dot{M}_\mathrm{wind}v_\mathrm{esc}^2$. In the APSW scenario it is supposed that the wind gets the
driving power from the energy deposited by accretion onto the stellar surface. The total available accretion power can be defined as \citep{matt05b}:
\begin{equation}
\dot{E}_\mathrm{acc}=\frac{1}{2}\dot{M}_\mathrm{acc}v_\mathrm{esc}^2\left[1-\frac{R_\star}{2R_t}-\delta\left(\frac{R_t}{R_\star}\right)^{1/2}\right] \; ,
\label{eq:lacc}
\end{equation}
given by the sum of the change in potential energy and kinetic energy minus the work done by the accretion torque. Terms proportional to $\delta^2$ have
been neglected. If we assume that the stellar wind consumes a fraction 
$f_\mathrm{E}=\dot{E}_\mathrm{wind}/\dot{E}_\mathrm{acc}$ of the accretion power, the remaining accretion shock luminosity
$L_\mathrm{UV}=\dot{E}_\mathrm{acc}-\dot{E}_\mathrm{wind}$ is given by:
\begin{eqnarray}
L_\mathrm{UV}=\frac{1}{2}\dot{M}_\mathrm{acc}v_\mathrm{esc}^2\left[1-0.5C^{-1}\Psi^{-2/7}-\delta C^{1/2}\Psi^{1/7}\right.\nonumber\\
\left.-\left(\frac{f_\mathrm{J}C^{1/2}\Psi^{1/7-2m}}{K^2\delta}\right)^\frac{1}{1-2m}\right]\; .
\label{eq:luv}
\end{eqnarray}
The accretion luminosity is usually employed to estimate the accretion rate assuming that all the accretion power is radiated at the 
accretion shock. This translates into a simplified formula like $\dot{M}_\mathrm{obs}=k\,L_\mathrm{UV}R_\star/GM_\star$, with $k$ a numerical factor of 
order unity \citep[e.g. $k=1.25$, ][]{Gullbring98}, reflecting the uncertainty on the position of the truncation radius. This determines a possible discrepancy
between the observed accretion rate $\dot{M}_\mathrm{obs}$ and the real one $\dot{M}_\mathrm{acc}$.

We try now to apply this simple model to some CTTS observations.
We selected a sample of stars (see Table~\ref{table:sample}) for which spectropolarimetric observations are available, so that a magnetic topology reconstruction is possible: 
V2129~Oph \citep{Donati07,Donati10b},  BP~Tau \citep{Donati08}, CV~Cha, CR~Cha \citep{Hussain09} and AA~Tau \citep{Donati10a}. 
In the case of V2129~Oph, BP~Tau and AA~Tau the measurements of the dipolar component are available; in the case of CV~Cha and CR~Cha, we take maximum observed 
magnetic field as the maximum of the dipolar component, since the authors did not provide the intensity of the multipoles.  
In Table~\ref{table:sample} we report the intensity of the dipolar field at the stellar equator, which corresponds to half of its maximum, located at the pole.
For V2129~Oph, BP~Tau and AA~Tau two values 
of the accretion luminosity are provided: Donati and collaborators have recently re-estimated this quantity, giving luminosities which are systematically around one order 
of magnitude smaller than previous estimates \citep[see e.g.][for BP~Tau and AA~Tau]{Gullbring98}. 
We also take into account a Taurus-Auriga sample from \citet{JK07}: for these stars the average magnetic field
has been measured ($\sim$ 1-3 kG), but the actual intensity of their dipolar component is unknown.

\section{An illustrative case: BP~Tau}
\label{sec:BP}

Taking as an example the specific case of BP~Tau, we assume the following values: $M_\star=0.7\;M_\odot$, $R_\star=1.95\;R_\odot$, $B_\star=600$ G, $P_\star=7.6$ d, 
$\delta=0.05$ \citep{Donati08}. Fixing these quantities, the accretion luminosity $L_\mathrm{UV}$ depends only on the parameters $\dot{M}_\mathrm{acc}$ and $f_\mathrm{J}$. 
In  Fig.~\ref{fig:bptau}  (left panel), we plot the accretion luminosity as a function of the ``true'' accretion rate (not the ``observed'' one) for three different values of $f_\mathrm{J}$ 
(Eq.~\ref {eq:luv}). The difference between these curves and the available accretion power (dashed line) gives the power consumed to drive the stellar wind: since the energy 
requirement of the wind $f_\mathrm{E}$ increases with the accretion rate, it is possible to find a maximum accretion luminosity consistent with a given $f_\mathrm{J}$ value.
Therefore, in the BP~Tau case, a spin equilibrium situation ($f_\mathrm{J}=1$) is only compatible with an accretion luminosity smaller than $L_\mathrm{UV}\simeq0.027L_\odot$. 
This is actually consistent with the recent estimate $L_\mathrm{UV}=0.023L_\odot$ \citep{Donati10a}, which in fact allows a maximum spin-down efficiency 
$f_\mathrm{J}=1.06$. However, the higher luminosity ($L_\mathrm{UV}=0.179L_\odot$)
estimated by \citet{Gullbring98} is not compatible with a zero-torque condition, independently of the value of the accretion rate and it allows a 
maximum spin-down efficiency $f_\mathrm{J}=0.52$: the stellar wind is consuming too much accretion power. Anyway, it is possible to see in the right panel of Fig.~\ref{fig:bptau}
that, independently of the accretion luminosity, the properties of a stellar wind at maximum spin-down efficiency are very demanding, having a mass flux $f_\mathrm{M}\sim0.4-0.5$ 
and consuming an important fraction of the accretion power ($f_\mathrm{E}\sim0.6$). For $f_\mathrm{J}<0.2$ the wind characteristics are less tough.
 
A summary of the limits obtained applying the APSW model to the star sample of Table~\ref{table:sample} is given in Table~\ref{table:limits}. 
For each star ($M_\star,R_\star,B_\star,P_\star,L_\mathrm{UV}$ specified) we show the maximum spin-down efficiency $f_\mathrm{J}$ and the corresponding mass outflow 
rate $f_\mathrm{M}$, energy requirement $f_\mathrm{E}$, accretion rate and wind outflow rate. If a value $f_\mathrm{J}>1$ is found, we show in parentheses
 the characteristics of the same star-wind system at spin equilibrium ($f_\mathrm{J}=1$).

\begin{deluxetable*}{c  l c c c c c c c c c}
\tablecaption{Limiting parameters of the stellar wind\label{table:limits}}
\tablewidth{0pt}
\tablehead{
\multicolumn{2}{c}{Object}  & \colhead{$f_\mathrm{J}$} & \colhead{$f_\mathrm{M}$} & \colhead{$f_\mathrm{E}$} &  \colhead{$\dot{M}_\mathrm{acc}$ (max)} & \colhead{$\dot{M}_\mathrm{wind}$ (max)} \\        
\multicolumn{2}{c}{Name} & \colhead{(max)} &  \colhead{(max)} & \colhead{(max)} & \colhead{[$M_\odot \;  \mathrm{\mbox{yr}}^{-1}$]} & \colhead{[$M_\odot \;  \mathrm{\mbox{yr}}^{-1}$]} \\
}
\startdata
\multirow{2}{*}{BP~Tau}  & (a)     & 0.52       & 0.42 & 0.6  & $5.5 \times 10^{-8}$ & $2.9 \times 10^{-8}$ \\
                                           & (b)     & 1.06 (1) & 0.49 (0.36) & 0.63 (0.56) & 7 (4.7)$\times 10^{-9}$ & 3.4 (1.7)$\times 10^{-9}$ \\
\multirow{2}{*}{V2129~Oph} & (a) & 0.31 & 0.32 & 0.55 & $3 \times 10^{-8}$ & $9.6\times 10^{-9}$ \\
                                                  & (b) & 1.24 (1) & 0.5 (0.24) & 0.63 (0.3) & 1.7 (0.9)$\times 10^{-9}$ & 8.8 (2.1)$\times 10^{-10}$ \\
CV~Cha  &    & 0.36 & 0.29 & 0.54 & $ 9.5 \times 10^{-8}$ & $2.8\times 10^{-8}$ \\
CR~Cha  &    & 1.7 (1) &  0.39 (0.09) & 0.63 (0.16) &  3.8 (1.7)$\times 10^{-9}$ & 1.3 (0.15)$\times 10^{-9}$\\
\multirow{2}{*}{AA~Tau} & (a) &  1.8 (1) & 0.52 (0.11) & 0.64 (0.14) &  7.9 (3.3)$\times 10^{-9}$ & 4.1 (0.36)$\times 10^{-9}$\\
                                           & (b) &  2.8 (1) & 0.52 (0.05) & 0.65 (0.06) &  1.8 (0.7)$\times 10^{-9}$ & 9.5 (0.32)$\times 10^{-10}$\\
\enddata
\end{deluxetable*}

\section{Spin equilibrium}
\label{sec:spin}

 \begin{figure}[t!]
  \includegraphics[width=\columnwidth]{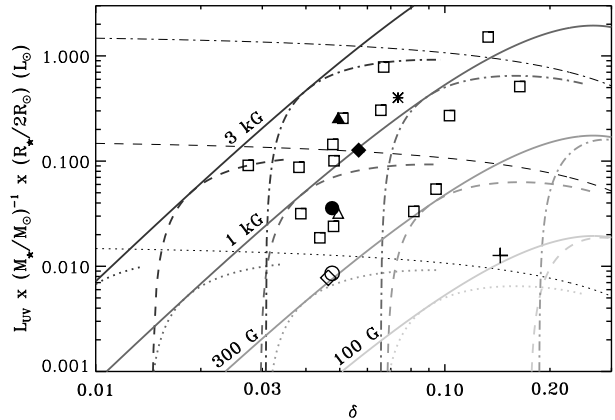}
  \caption{Accretion luminosity $L_\mathrm{UV}$ - stellar spin $\delta$ solutions compatible with the presence of an APSW at spin equilibrium. Linestyles 
                  correspond to different accretion rates: $\dot{M}_\mathrm{acc}=10^{-9}\;M_\odot\;\mathrm{\mbox{yr}}^{-1}$ (dotted lines), 
                  $\dot{M}_\mathrm{acc}=10^{-8}\;M_\odot\;\mathrm{\mbox{yr}}^{-1}$  (dashed lines), 
                  $\dot{M}_\mathrm{acc}=10^{-7}\;M_\odot\;\mathrm{\mbox{yr}}^{-1}$ (dot-dashed lines). 
                   Shades of grey correspond to different normalized values of the 
                   stellar magnetic field $B_\star\left(M_\star/M_\odot\right)^{-1/4}\left(R_\star/2R_\odot\right)^{5/4}$. Solid lines represent the envelopes of the 
                   spin equilibrium solutions for a given magnetic field. Black thin lines with different linestyles correspond to the maximum accretion power available for different 
                   accretion rates (Eq.~\ref{eq:lacc} with $R_\mathrm{t}=R_\mathrm{co}$). Symbols correspond to observations of BP~Tau (a and b, filled and empty triangles), 
                   V2129~Oph (a and b, filled and empty diamonds), AA~Tau (a and b, filled and empty circles), CV~Cha (asterisk), CR~Cha (plus), \cite{JK07} 
                   sample (empty squares). }
  \label{fig:LD}
\end{figure}

In Fig.~\ref{fig:LD} we plot several $L_\mathrm{UV}-\delta$ solutions of Eq.~(\ref{eq:luv}) at spin equilibrium ($f_J=1$) for different 
$B_\star \left(M_\star/M_\odot\right)^{-1/4}\left(R_\star/2R_\odot\right)^{5/4}$ values (using different shades of grey) and different $\dot{M}_\mathrm{acc}$ values 
(using different linestyles). We also truncate the solutions whenever the truncation radius is equal to the corotation radius:
$R_\mathrm{t}=R_\mathrm{co}=\delta^{-2/3}R_\star$. In fact, when the disk truncation occurs very close to corotation, 
disk-locked solutions are in principle possible \citep{matt05a}, while, for $R_\mathrm{t}>R_\mathrm{co}$, accretion 
becomes intermittent and highly variable \citep[``propeller'' regime,][]{Ustyu06}: in both cases  the magnetic star-disk interaction can provide an 
efficient enough braking torque.

The figure shows that, for a given magnetic field $B_\star$, the accretion luminosity - $\delta$ configurations at spin equilibrium have an envelope
depending on the normalized $B_\star$ value (solid lines): the solid curves give therefore an estimate of the minimum dipolar component compatible with 
a spin equilibrium condition.
Applying the model to our entire star sample, it is possible to see that dipolar fields between 100 G and 3 kG are
required to power an APSW at spin equilibrium compatible with the observed accretion luminosity. In many cases
a field stronger than 1~kG is required, which has been currently observed only in the case of AA~Tau.

By fitting the solid curves in Fig.~\ref{fig:LD} we can define the minimum intensity of the dipolar component required by a spin equilibrium configuration, given the 
characteristic of a particular star ($L_\mathrm{UV}$, $M_\star$, $R_\star$, $\delta$):
\begin{eqnarray}
B_{\star,\mathrm{min}} && \approx1025\left(\frac{L_\mathrm{UV}}{0.1L_\odot} \right)^{1/2}\left(\frac{M_\star}{M_\odot}\right)^{-1/4}\left(\frac{R_\star}{2R_\odot}\right)^{-3/4}\nonumber\\
                    \times    &&  \left[0.933\left(\frac{\delta}{0.05}\right)^{-1.5}+0.067\left(\frac{\delta}{0.05}\right)^{0.45}\right]\;\mathrm{G}\;.
\label{eq:blim}                           
\end{eqnarray}
Notice that Eq.~(\ref{eq:blim}) gives a lower limit for the dipolar field intensity at the stellar {\it equator}, while the value at the stellar {\it pole} is twice larger. 
The dependency on $\delta$ has been obtained by fitting the solutions in the range $0.01< \delta <0.266$, where $\delta \approx 0.266$ corresponds to an extrema of the solid
curves in Fig.~\ref{fig:LD}. 
The other dependencies are exact.  We can also estimate the mass and energy efficiencies $f_M$ and $f_E$ of the stellar wind which correspond to 
the solutions of Eq.~(\ref{eq:blim}):
\begin{eqnarray}
f_{M,\mathrm{max}} & \approx & 0.48\left[1.55 -0.55\left(\frac{\delta}{0.05}\right)^{0.467}\right]\nonumber \\
f_{E,\mathrm{max}}  & \approx &  0.63\left[1.03-0.03\left(\frac{\delta}{0.05}\right)^{1.04}\right]\;.
\label{eq:efflim}
\end{eqnarray}
These values represent the maximum efficiencies required by the spin equilibrium: for a dipolar component stronger than the intensity defined by Eq.~(\ref{eq:blim}), 
both efficiencies can be lower (see for example the values between parentheses in Table \ref{table:limits}). On the other hand, a stronger field determines at some point 
the transition to a propeller regime ($R_\mathrm{t}>R_\mathrm{co}$). Imposing $\Psi=(\delta^{-2/3}/C)^{7/2}$ in Eq.~(\ref{eq:ejeff}) and (\ref{eq:lacc}), we can therefore 
estimate the minimum ejection and energy efficiencies of a stellar wind at spin equilibrium in a non-propeller regime:
\begin{eqnarray}
\label{eq:fmmin}
f_\mathrm{M,min} & \approx & 0.15\left(\frac{\delta}{0.05}\right)^{-0.53}\\
f_\mathrm{E,min}  & \approx & 0.18\left(\frac{\delta}{0.05}\right)^{-0.53}\left[1.25 -0.25\left(\frac{\delta}{0.05}\right)^{2/3}\right]^{-1}\nonumber\;.  
\end{eqnarray}
Anyway, the upper limits on $f_\mathrm{M}$ and $f_\mathrm{E}$ are likely more robust than the lower ones. When the truncation radius approaches
corotation, which is the condition used to derive Eq. (\ref{eq:fmmin}), it is not certain if our approximation for the truncation radius is still valid. Using the $C$ factor
as a measure of this uncertainty, the $f_\mathrm{M,min}$ value is more sensible to errors on the truncation position ($f_\mathrm{M,min}\propto C^{2.82}$) than 
$f_\mathrm{M,max}$ ($f_\mathrm{M,max}\propto C^{0.22}$ around the fiducial values $C=0.75$, $\delta=0.05$).
 A field larger than the Eq.~(\ref{eq:blim}) limit would be also compatible with a 
$f_\mathrm{J}>1$ situation, but it would require mass and energy efficiencies even greater than the Eq.~(\ref{eq:efflim}) estimates (see Table  \ref{table:limits}).

\section{Discussion and conclusions}
\label{sec:dicsussion}

We applied the accretion powered stellar wind model \citep{matt05b} to a sample of CTTS to verify if stellar winds are a viable mechanism to spin-down 
the rotation of accreting and contracting protostars. According to this scenario, a fraction of the energy deposited by the magnetospheric accretion flow onto the surface 
of the star could be used to drive the stellar wind: we added the additional constraint that the same accretion energy must be used to power the emission 
of the accretion shock. In Section \ref{sec:BP} we showed that, for a given spin-down efficiency $f_\mathrm{J}\neq0$, 
a maximum accretion luminosity can be attained: when the accretion power and, consequently, the spin-up torque become too large, the stellar wind consumes too much
energy and a smaller and smaller fraction is left to support the emission.  

In Table \ref{table:limits} we showed that the stars in our sample characterized by a high accretion luminosity ($L_\mathrm{UV}\geq0.1L_\odot$) impose severe limits on 
the accretion power which is available to drive the stellar wind so that the spin-down torque is not strong enough to achieve a spin equilibrium. 
This is consistent with Eq.~(\ref{eq:blim}): in the range of parameters covered in our sample, a dipolar component of kG intensity is required by the
wind to spin-down a star characterized by such a high UV emission.  In Fig.~\ref{fig:LD} we also showed that an important fraction of our sample (around $50\%$) would require
such a strong dipolar component to be compatible with a zero-torque condition: at the moment of writing a dipolar component of kG intensity has been measured
only in the case of AA~Tau. 

Equation (\ref{eq:blim}) and Fig.~\ref{fig:LD} clearly show that lower UV luminosity and/or faster spinning stars require weaker fields, more consistent with the dipolar 
intensities currently measured, to be in spin equilibrium with an APSW. In fact, stars in the sample with a low accretion luminosity ($L_\mathrm{UV}\ll0.1L_\odot$) are compatible with a 
$f_\mathrm{J}\geq1$ situation, but the corresponding APSW at maximum spin-down efficiency is energetically very 
demanding, as confirmed by  Eq.~(\ref{eq:efflim}). Some low luminosity cases at spin equilibrium are less demanding, see e.g.  the CR~Cha or AA~Tau examples. Still, 
the mass fluxes would correspond roughly to the entire mass flux of T~Tauri jets \citep[1-20$\%$,][]{Cabrit09}: this would imply that stellar winds are the primary 
ejecting component of young stars, which seems unlikely \citep{Ferreira06}.
Besides, even for relatively low ejection rates ($f_\mathrm{M}\leq0.1$) the energy input is still an issue. It is already  known that the wind can not be thermally driven 
\citep{matt07}: a temperature close to virial ($\sim10^6$ K) determines a too high emission and is incompatible with observations \citep{Dupree05}. 
Turbulent Alfv\'en waves represent another possible pressure 
source \citep{Dec81}. Furthermore, it has been suggested that the amplitude of the waves generated by the impact of the accretion streams onto the surface of the star
is greater than interior convection-driven wave amplitude \citep{Cranmer09}. In this case, the accretion/ejection energy coupling is not easy to determine:
recent models suggest anyway that the wind mass loss rates due to this mechanism are generally very low \citep[$10^{-5}<f_\mathrm{M}<10^{-2}$,][]{Cranmer09}.
Besides, it is important to remark that in our sample, when the APSW ejection efficiency $f_\mathrm{M}$ at spin equilibrium becomes $\leq 0.1$, the star-disk system 
approaches a propeller regime ($R_\mathrm{t}\gtrsim R_\mathrm{co}$, see Eq.~\ref{eq:fmmin}), as in the typical AA~Tau case \citep{Donati10a}. 
When $R_\mathrm{t}\gtrsim R_\mathrm{co}$, the spin-down torque due to the star-disk interaction, which has not been taken into account here, is in principle 
enough to slow down the stellar rotation \citep{matt05a, Ustyu06}.

We therefore conclude that accretion powered stellar winds are unlikely to be the sole mechanism to provide an efficient spin-down torque for accreting classical 
T~Tauri stars. Our study suggests that a conservative limit  on the wind spin-down torque ($f_\mathrm{J}<0.1-0.2$)  reduces the mass flux and power requirements 
to values more compatible with models of wave-driven winds from T~Tauri stars \citep[$f_\mathrm{M}\sim f_\mathrm{E}<1\%$,][]{Cranmer08}. 
The problem of the spin of accreting and contracting stars like T~Tauri has still many  open issues. It is likely that diverse mechanisms contribute at the same time with 
different degrees: stellar winds, magnetospheric star-disk angular momentum exchanges \citep{ZF09}, magnetospheric ejections driven by the 
star-disk interaction \citep{Ferreira00,Zanni09}. 

\acknowledgments

CZ acknowledges support from the Marie Curie Action ``European Reintegration Grants'' under contract PERG05-GA-2009-247415.
We thank the referee, Sean Matt, for his helpful comments.


\begin{thebibliography}{}
\bibitem[Bessolaz et al.(2008)]{Bessolaz08} Bessolaz, N., Zanni, C., Ferreira, J., Keppens, R. \& Bouvier, J. 2008, \aap, 478, 155
\bibitem[Bouvier et al.(1993)]{Bouvier93} Bouvier, J., Cabrit, S., Fern\'andez, M., Mart\'in, E. L. \& Matthews, J. M. 1993, \aap, 272, 176
\bibitem[Cabrit(2009)]{Cabrit09} Cabrit, S. 2009, in Protostellar Jets in Context, ed. K. Tsinganos, T. Ray \& M. Stute (Berlin: Springer), 247
\bibitem[Cranmer(2008)]{Cranmer08} Cranmer, S. R. 2008, \apj, 689, 316
\bibitem[Cranmer(2009)]{Cranmer09} Cranmer, S. R. 2009, \apj, 706, 824
\bibitem[DeCampli(1981)]{Dec81} DeCampli, W. M. 1981, \apj, 244, 124
\bibitem[Donati et al.(2007)]{Donati07} Donati., J.-F., Jardine, M. M., Gregory, S. G. et al. 2007, \mnras, 380, 1297
\bibitem[Donati et al.(2008)]{Donati08} Donati., J.-F., Jardine, M. M., Gregory, S. G. et al. 2008, \mnras, 386, 1234
\bibitem[Donati et al.(2010a)]{Donati10a} Donati, J.-F., Skelly, M. B., Bouvier, J. et al. 2010a, \mnras, 409, 1347
\bibitem[Donati et al.(2010b)]{Donati10b} Donati, J.-F., Bouvier, J., Walter, F. M. et al. 2010b, \mnras, in press (\texttt{arXiv:1011.4789})
\bibitem[Dupree et al.(2005)]{Dupree05} Dupree, A. K., Brickhouse, N. S., Smith, G. H. \& Strader, J. 2005, \apj, 625, L131
\bibitem[Ferreira et al.(2000)]{Ferreira00} Ferreira, J., Pelletier, G. \& Appl, S. 2000, \mnras, 312, 387
\bibitem[Ferreira et al.(2006)]{Ferreira06} Ferreira, J., Dougados, C. \& Cabrit, S. 2006, \aap, 453, 785
\bibitem[Ghosh \& Lamb(1979)]{GL79} Ghosh P. \& Lamb F. K. 1979, \apj, 234, 296
\bibitem[Gullbring et al.(1998)]{Gullbring98} Gullbring, E., Hartmann, L., Briceno, C. \& Calvet N. 1998, \apj, 492, 323
\bibitem[Gullbring et al.(2000)]{Gull00} Gullbring, E., Calvet, N., Muzerolle, J. \& Hartmann L. 2000, \apj, 544, 927
\bibitem[Herbst et al.(2007)]{Herbst07} Herbst ,W., Eisl\"offel, J.,Mundt, R. \& Scholz, A. 2007, in Protostars and Planets V, 
                                                                     ed. B. Reipurth, D. Jewitt \& K. Keil (Tucson, AZ: Univ. Arizona Press), 297
\bibitem[Hussain et al.(2009)]{Hussain09} Hussain, G. A. J., Collier Cameron, A., Jardine, M. M. et al. 2009, \mnras, 398, 189
\bibitem[Irwin \& Bouvier(2009)]{Irwin09} Irwin, J. \& Bouvier, J. 2009, in IAU Symp. 258, The Ages of Stars, ed. E. E.
                                                  Mamajek, D. R. Soderblom, \& R. F. G.Wyse (Cambridge: Cambridge Univ. Press), 363
\bibitem[Johns-Krull(2007)]{JK07} Johns-Krull, C. M. 2007, \apj, 664, 975
\bibitem[K\"onigl(1991)]{ko91} K\"onigl, A.  1991, \apj, 370, L39
\bibitem[Lamm et al.(2005)]{Lamm05} Lamm, M. H., Mundt, R., Bailer-Jones, C. A. L. \& Herbst, W. 2005, \aap, 430, 1005
\bibitem[Matt \& Pudritz(2005a)]{matt05a} Matt, S. \& Pudritz, R. E. 2005a, \mnras, 356, 167 
\bibitem[Matt \& Pudritz(2005b)]{matt05b} Matt, S. \& Pudritz, R. E. 2005b, \apj, 632, L135
\bibitem[Matt \& Putritz(2007)]{matt07} Matt, S. \& Pudritz, R. E. 2007, in IAU Symposium, Vol. 243, Star-Disk Interaction in Young Stars, 
            ed. J. Bouvier \& I. Appenzeller, 299
\bibitem[Matt \& Pudritz(2008a)]{Matt08a} Matt, S. \& Pudritz, R. E. 2008a, \apj, 678, 1109
\bibitem[Matt \& Pudritz(2008b)]{Matt08b} Matt, S. \& Pudritz, R. E. 2008b, \apj, 681, 391
\bibitem[Romanova et al.(2009)]{Romanova09} Romanova, M. M., Ustyugova, G. V., Koldoba, A. V. \& Lovelace, R. V. E. 2009, \mnras, 399, 1802
\bibitem[Scheurwater \& Kuijpers(1988)]{Sch88} Scheurwater, R. \& Kuijpers, J. 1988, \aap, 190, 178
\bibitem[Shu et al.(1994)]{Shu94} Shu, F. H., Najita, J., Ostriker, et al. 1994, \apj, 429, 781
\bibitem[Ustyugova et al.(2006)]{Ustyu06} Ustyugova, G. V., Koldoba, A. V., Romanova, M. M. \& Lovelace, R. V. E. 2006, \apj, 646, 304
\bibitem[Zanni(2009)]{Zanni09} Zanni, C. 2009, in Protostellar Jets in Context, ed. K. Tsinganos, T. Ray \& M. Stute (Berlin: Springer), 165
\bibitem[Zanni \& Ferreira(2009)]{ZF09} Zanni, C. \& Ferreira, J. 2009, \aap, 508, 1117
\end{thebibliography}
\end{document}